\documentclass[conference]{IEEEtran}
\usepackage{cite}
\usepackage[utf8]{inputenc} 
\usepackage[T1]{fontenc}
\usepackage{amssymb,amsfonts}
\usepackage[cmex10]{amsmath}
\usepackage{algorithm}
\usepackage{algpseudocode}
\let\oldReturn\Return
\renewcommand{\Return}{\State\oldReturn}
\allowdisplaybreaks
\usepackage{graphicx}
\usepackage{textcomp}
\usepackage{xcolor}
\usepackage{pgfplots}
\usepackage{subcaption}
\usepackage[keeplastbox]{flushend}
\usepackage{tikz}

\usetikzlibrary{matrix, positioning, arrows}

\newtheorem{theorem}{Theorem}

\newtheorem{lemma}{Lemma}

\def\BibTeX{{\rm B\kern-.05em{\sc i\kern-.025em b}\kern-.08em
    T\kern-.1667em\lower.7ex\hbox{E}\kern-.125emX}}

\begin{document}

\title{A New Permutation Decoding Method for Reed-Muller Codes
}

\author{
\IEEEauthorblockN{Mikhail Kamenev, Yulia Kameneva, Oleg Kurmaev, Alexey Maevskiy}
\IEEEauthorblockA{Moscow Research Center, Huawei Technologies Co., Ltd. \\
Moscow, Russia \\
Email: \{kamenev.mikhail1, kameneva.iuliia, oleg.kurmaev, maevskiy.alexey\}@huawei.com}
}

\maketitle

\begin{abstract}
A novel permutation decoding method for Reed-Muller codes is presented. The complexity and the error correction performance of the suggested permutation decoding approach are similar to that of the recursive lists decoder. It is demonstrated that the proposed decoding technique can take advantage of several early termination methods leading to a significant reduction of the operations number required for the decoding, with the error correction performance being the same.
\end{abstract}

\section{Introduction}
Reed-Muller (RM) codes are a family of error correcting codes discovered by Muller \cite{Muller} and shortly after by Reed \cite{Reed}, who also proposed the first efficient decoding algorithm. Recently it has been proven that RM codes achieve the capacity on an erasure channel under maximum a posteriori (MAP) decoding \cite{Urbanke}. Unfortunately, practical usage of MAP decoding is limited by its complexity. If it is not feasible to use MAP decoding of RM codes, then sub-optimal algorithms, i.e. a recursive lists decoder \cite{Dumer}, can be used with a degradation of the error correction performance of the code.  

RM codes may be considered as polar codes with the appropriate selection of the frozen bits set \cite{Polar}. Polar codes have been shown to achieve the symmetric capacity of any binary-input discrete memoryless channel under a low-complex successive cancellation (SC) decoder \cite{Polar}. However, the performance of finite length polar codes under the SC decoder is quite poor. A successive cancellation list (SCL) decoder allows getting performance very close to that of maximum-likelihood decoding \cite{SCL}. Observe that SCL decoding is similar to the recursive lists algorithm. Here we will consider RM codes from polar codes point of view.

In the paper, a new permutation decoding method for RM codes is proposed. This decoder has the complexity similar to that of the SCL decoder, namely $\mathcal{O}(Ln\log{}n)$, where $n$ is the code length and $L$ is the list size. In contrast with the SCL decoder, it does not use sorting operation, which is challenging for the hardware implementation.
It also benefits from several early termination techniques, significantly decreasing the number of calculations in comparison with the SCL algorithm. The error correction performance of the considered decoder is similar to that of the SCL decoding. Moreover, a parallel implementation of the proposed decoder is possible, leading to the decoding latency $\mathcal{O}(n\log{}n)$.  

The rest of the paper is organized as follows. Section II provides a general description of RM codes, polar codes, and its decoding algorithm. In section III a new permutation decoding method for RM codes is presented. In section IV we propose three early termination methods for the proposed decoder. Numerical results are presented in section V. We conclude the paper in section VI.

\section{RM and polar codes}
$(n, k)$ polar code \cite{Polar} is a linear block code of length $n=2^m$, where $m$ is some positive integer, and dimension $k$ generated by $k$ rows $j_i \in \left\lbrace0, 1, \dots , n - 1\right\rbrace \setminus \mathcal{F}$, $0 \leq i < k$ of the matrix
\begin{equation}
\mathbf{A_m} = \begin{bmatrix}
    1 & 0 \\
    1 & 1 \\
\end{bmatrix}^{\otimes m},
\end{equation}
where $\mathbf{X}^{\otimes m}$ denotes $m$-times Kronecker product of the matrix $\mathbf{X}$ with itself. The set of frozen bits $\mathcal{F}$ is constructed as a set of indices $i$ maximizing  error correction performance of the code. For instance, Gaussian approximation (GA) for density evolution \cite{GA} can be used to generate polar codes having optimal error correction performance under the SC decoding algorithm in the binary-input additive white Gaussian noise (BI-AWGN) channel. 

RM code with parameters $r$ and $m$ is a linear block code of length $n=2^m$ and dimension $k$ generated by $k$ rows $\mathbf{r_0}, \mathbf{r_1}, \dots \mathbf{r_{k-1}}$ of the matrix $\mathbf{A_m}$ such that $\lVert \mathbf{r_{i}}\rVert \geq 2^{m - r}$, where $\lVert \mathbf{x}\rVert$ denotes Hamming weight. Since RM codes have the same form of the generator matrix, they can be constructed as polar codes with the specific choice of the frozen bits set.

%Polar codes Log Likelihood Ratio (LLR) based SC decoding procedures can be efficiently implemented using the factor graph representation \cite{Polar}. The factor graph of polar code of length $2^m$ contains $m$ layers. Let $y_{i}^{m+1}$ denotes channel LLR with index $i$, while $y_{i}^{l}$ denotes LLR with index $i$, after proccessing by the layer $l$. Then the decoder is based on the following operations
%\begin{subequations}
%\begin{align}
%y_{i}^{l-1} &= f_{-}\left(y_{i}^{l}, y_{i + 2^{l - 1}}^{l}\right), \label{xorY} \\ 
%y_{i+2^{l - 1}}^{l-1} &= f_{+}\left(y_{i}^{l}, y_{i + 2^{l - 1}}^{l}, \hat{u}_{i}^{l-1} \right), \label{sumY}
%\end{align}
%\end{subequations}
%where $i \in \bigcup\limits_{g = 0}^{2^{m - l}}\left\lbrace 2^{l}g, \dots, 2^{l}g + 2^{l-1}-1\right\rbrace.$ $f_{-}$ and $f_{+}$ are defined as
%\begin{subequations}
%\begin{align}
%f_{-}(x,y) &\triangleq \ln_{}\left(\frac{e^{x + y} + 1}{e^{x} + e^{y}}\right), \label{softXor} \\ 
%f_{+}(x,y,u) &\triangleq \left(1 - 2u\right)x + y,  \label{softSum}
%\end{align}
%\end{subequations}
%where $x$ and $y$ are real, while $u$ is binary. Instead of using $f_{-}$, we will follow the approach proposed in \cite{MinSum}, and use the hardware friendly approximation, namely
%\begin{equation}
%f_{-}\left(x,y\right) \approx \tilde{f_{-}}\left(x,y\right)\triangleq \textrm{sign}\left(x\right)\textrm{sign}\left(y\right)\textrm{min}\left(\left|x\right|, \left|y\right|\right)
%\end{equation}

Polar codes encoding and decoding procedures can be efficiently implemented using the factor graph representation \cite{Polar}. The factor graph of a code of length $2^m$ contains $m$ layers of operations. Let $u_{i}^{0}$ and $u_{i}^{m}$ denotes information bit and codeword bit respectively. The subscript denotes bit index. Then information bits are processed layer by layer using following update rules: 
\begin{equation}
\begin{gathered}
u_{i}^{l+1} = u_{i}^{l} \oplus u_{i + 2^l}^{l}, \\
u_{i + 2^l}^{l+1} = u_{i + 2^l}^{l}, \\
i \in \bigcup\limits_{g = 0}^{2^{m - l - 1}}\left\lbrace 2^{l+1}g,2^{l+1}g + 1, \dots, 2^{l+1}g + 2^{l}-1\right\rbrace.
\end{gathered}
\label{EncodingRec}
\end{equation}
 The same factor graph is used for log likelihood ratio (LLR) based SC decoding \cite{MinSum}. Channel LLRs $y_{i}^{m}$ are processed in a recursive manner, namely 
\begin{subequations}
\begin{align}
y_{i}^{l-1} &= f_{-}\left(y_{i}^{l}, y_{i + 2^{l - 1}}^{l}\right), \label{xorY} \\ 
y_{i+2^{l - 1}}^{l-1} &= f_{+}\left(y_{i}^{l}, y_{i + 2^{l - 1}}^{l}, \hat{u}_{i}^{l-1} \right), \label{sumY}
\end{align}
\end{subequations}
where $y_{i}^{0}$ and $y_{i}^{m}$ denotes LLRs used for the information bits evaluation and LLRs received from a channel respectively, while the subscript denotes LLR index, $i \in \bigcup\limits_{g = 0}^{2^{m - l}}\left\lbrace 2^{l}g, 2^{l}g + 1, \dots, 2^{l}g + 2^{l-1}-1\right\rbrace$. $\hat{u}_{i}^{j}$ is a bit value calculated by the SC algorithm using (\ref{EncodingRec}). $f_{-}$ and $f_{+}$ are defined as
\begin{subequations}
\begin{align}
f_{-}(x,y) &\triangleq \ln_{}\left(\frac{e^{x + y} + 1}{e^{x} + e^{y}}\right), \label{softXor} \\ 
f_{+}(x,y,u) &\triangleq \left(1 - 2u\right)x + y,  \label{softSum}
\end{align}
\end{subequations}
where $x, y \in \mathbb{R}$, $u \in \left\lbrace0,1\right\rbrace$. We will follow the approach proposed in \cite{MinSum} and use the hardware-friendly approximation of (\ref{softXor}), namely
\begin{equation}
f_{-}\left(x,y\right) \approx \tilde{f_{-}}\left(x,y\right)\triangleq \textrm{sign}\left(x\right)\textrm{sign}\left(y\right)\textrm{min}\left\lbrace\left|x\right|, \left|y\right|\right\rbrace.
\end{equation}

\section{Permutation decoder for RM codes}
Unlike polar codes, RM codes have the permutation group which is isomorphic to the whole affine group $GA(m)$ \cite[Sec.~13.9]{Sloan}. Recall that the permutation group of a code contains permutations of the code positions that does not change the set of codewords, i.e. transform any codeword of the code to another or the same codeword.
For simplicity, we will consider only $m!$ factor graph layers permutations $\pi^l:\left(0,1, \dots, m-1 \right) \rightarrow \left(\pi^l\left(0\right), \pi^l\left(1\right), \dots, \pi^l\left(m-1\right) \right)$ \cite{Perm}. Let $\pi$ be the corresponding bit indices permutation.

The suggested permutation decoding approach, as the SCL decoding algorithm, returns a list of $L$ codewords and then some metric is used to choose the best one. The LLR based metric considered in \cite{MinSum} is used, namely

\begin{equation}
M = \sum_{i \in \mathcal{F}} \textrm{min}\left\lbrace0,\left(1 - 2\hat{u}_i^0\right)y_i^0 \right\rbrace,
\label{PathMetricSCPerm}
\end{equation}
where $\hat{u}_i^0$ and $y_i^0$ denote a bit estimation and an LLR value respectively. They are obtained by the SC decoder, using (\ref{xorY}) and (\ref{sumY}). The metric benefits from the following theorem.
\begin{theorem} Consider a polar code of length $n = 2^m$ with the frozen bits set $\mathcal{F}$. Let $y_{0}^{m}, y_{1}^{m}, \dots, y_{n-1}^{m}$ denote received channel LLRs. Then 
\begin{equation}
\sum_{i \in \mathcal{F}} \textrm{min}\left\lbrace0,\left(1 - 2\hat{u}_i^0\right)y_i^0 \right\rbrace = \sum\limits_{i = 0}^{n-1} \textrm{min}\left\lbrace0,\left(1 - 2\hat{u}_i^m\right)y_{i}^{m} \right\rbrace,
\label{PathMetric}
\end{equation}
where $\hat{u}_i^j$ and $y_i^0$ are obtained using (\ref{xorY}) and (\ref{sumY}) after the finish of the SC decoding procedure. $\hat{u}_i^0$ denotes a frozen bit value, while $\hat{u}_i^m$ is a codeword bit.
\label{TheoremMetric}
\end{theorem}

The proof of the theorem is given in the appendix.

Here the following permutation decoding approach is proposed. Firstly, $L$ random permutations are generated. Then, the SC algorithm process $L$ permuted versions of the received channel LLRs, and return $L$ decoded codewords with corresponding metrics. Finally, the codeword with the best metric is returned as the output of the algorithm.

Since computational complexity of the SC decoder equals $\mathcal{O}(n\log{}n)$ \cite{Polar}, the complexity of the considered permutation decoder equals that of $L$ SC decoders, namely $\mathcal{O}(Ln\log{}n)$. Unlike the SCL decoder, the presented permutation decoder can return less than $L$ unique codewords. However, simulation results demonstrate that the error correction performance of the proposed algorithm is similar to that of the SCL decoder for large list size. Moreover, this fact can be used to significantly decrease the number of calculations, with the error correction performance degradation being negligible.

Another benefit of the proposed method is that it does not use the sorting operation. Thus, it is more feasible for hardware implementation than the SCL algorithm. Moreover, all copies of the SC decoder can be run in parallel, leading to the decoding latency $\mathcal{O}(n\log{}n)$.

\section{Early termination methods}
In the section, several methods aiming to decrease the number of $f_+$ and $\tilde{f_{-}}$ operations in the considered permutation decoding algorithm are proposed. They benefit from the metric used, knowledge about signal-to-noise ratio (SNR) in a BI-AWGN channel, and the fact that several copies of the SC decoder can return the same codeword.

\subsection{Branch and bounds method}
The goal of the considered permutation decoding technique is to find a codeword with the best, i.e. the biggest, metric. If the instances of the SC decoder are running sequentially, then one can use knowledge of the best metric found so far. Note that the SC decoder process bits sequentially. Let $M_i$ denotes a value of metric (\ref{PathMetricSCPerm}) after processing $\hat{u}_0^0, \hat{u}_1^0, \dots \hat{u}_{i-1}^0$. Observe that if $i < j$, then $M_i \leq M_j$. So, it is possible to adopt the branch and bounds method, namely if the current metric has been already smaller than the best one found so far, then one can stop the decoding process under the current permutation without the error correction performance degradation.

\subsection{SNR based approach}
Although the previous early termination method benefits from its simplicity, it cannot be used for a parallel implementation of the permutation decoding algorithm. To decrease the decoder latency, one can run SC decoding for all permutations in parallel and then choose the codeword with the best metric. Thus, the first estimation of the best metric value will be obtained after the decoding is finished. The problem can be partially solved by dividing the permutations into groups and running the SC algorithm in parallel for each group, with groups being processed sequentially. However, the modification affects both the decoder latency and the early termination gain. 

 \begin{figure}[t]
 \begin{algorithmic}[1]
 \renewcommand{\algorithmicrequire}{\textbf{Input:}}
\renewcommand{\algorithmicensure}{\textbf{Output:}}
\Require A vector of LLRs $\mathbf{\hat{y}^l}$, a vector of bits $\mathbf{\hat{u}^0}$, a set of the frozen bits $\mathcal{F}$, an index of outer code $g$, a layer index $l$, a metric threshold  $M_t$.
\Ensure A current metric value $M$, a vector of bits $\mathbf{\hat{u}^{l}}$.  If $M < M_t$, $-\infty$ is returned.
\Function{SC}{$\mathbf{\hat{y}^l}, \mathbf{\hat{u}^0}, \mathcal{F}, g, l, M_t$}
\State Set $\mathbf{\hat{u}^l}$ to be all zeros vector of size $2^l$
 \If {$l = 0$}
 \If {$g \in \mathcal{F}$}
 \State $\mathbf{\hat{u}^0}[g] \gets 0$, $\mathbf{\hat{u}^l}[0]\gets 0$
 \State $M \gets \textrm{min}\left\lbrace0, \mathbf{\hat{y}^l}[0]\right\rbrace$
 \Else
 \If {$\mathbf{\hat{y}^l}[0] \leq 0$}
 \State $\mathbf{\hat{u}^0}[g]\gets 1$, $\mathbf{\hat{u}^l}[0]\gets 1$
 \Else
 \State $\mathbf{\hat{u}^0}[g] \gets 0$, $\mathbf{\hat{u}^l}[0]\gets 0$
 \EndIf
 \State  $M \gets 0$
 \EndIf
 \Return $M$, $\mathbf{\hat{u}^{l}}$
 \EndIf
 \State Set $\mathbf{\hat{y}^{l-1}}$ to be all zeros vector of size $2^{l-1}$\
 \For {$i = 0$ to $2^{l-1}-1$}
 \State $\mathbf{\hat{y}^{l-1}}[i] \gets \tilde{f_{-}}\left(\mathbf{\hat{y}^{l}}[i],\mathbf{\hat{y}^{l}}[i+2^{l-1}]\right)$
 \EndFor
 \State $M, \mathbf{\hat{u}^{l-1}}  \gets \texttt{SC}\left(\mathbf{\hat{y}^{l-1}},\mathbf{\hat{u}^0},\mathcal{F},  2g, l - 1,M_t\right)$
 \If {$M < M_t$}
 \Return $-\infty$, $\mathbf{\hat{u}^{l}}$
 \EndIf
 \For {$i = 0$ to $2^{l-1}-1$}
 \State $\mathbf{\hat{u}^{l}}[i] \gets \mathbf{\hat{u}^{l-1}}[i]$
 \State $\mathbf{\hat{y}^{l-1}}[i] \gets f_{+}\left(\mathbf{\hat{y}^{l}}[i],\mathbf{\hat{y}^{l}}[ i + 2^{l-1}], \mathbf{\hat{u}^{l-1}}[i] \right)$
 \EndFor
 \State $M^\prime,\mathbf{\hat{u}^{l-1}} \gets \texttt{SC}\left(\mathbf{\hat{y}^{l-1}},\mathbf{\hat{u}^0},\mathcal{F},  2g + 1, l - 1,M_t\right)$
 \State $M \gets M + M^\prime$
 \If {$M < M_t$}
 \Return $-\infty$, $\mathbf{\hat{u}^{l}}$
 \EndIf
 \For {$i = 0$ to $2^{l-1}-1$}
 \State $\mathbf{\hat{u}^{l}}[i] \gets \mathbf{\hat{u}^{l}}[i] \oplus \mathbf{\hat{u}^{l-1}}[i]$
 \State $\mathbf{\hat{u}^{l}}[i+ 2^{l-1}] \gets \mathbf{\hat{u}^{l-1}}[i]$
 \EndFor
 \Return $M$, $\mathbf{\hat{u}^{l}}$ 
 \EndFunction 
 \end{algorithmic} 
 \caption{Recursive calculations used in the SC algorithm.} 
 \label{RecursivelyCalcSC}
 \end{figure} 
  
The issue can be solved for a BI-AWGN channel using Theorem \ref{TheoremMetric} and information about the channel noise variance $\sigma^2$. Assume that all zeros codeword of length $n=2^m$ has been transmitted over a BI-AWGN channel with the noise variance $\sigma^2$,  using binary phase-shift keying modulation, and LLRs $y_0^m, y_1^m, \dots y_{n-1}^m$ are received. Then $y_i^m$ is sampled from a Gaussian random variable with mean $2/\sigma^2$ and variance $4/\sigma^2$ \cite[Sec.~7.3]{Sarah}. For simplicity, (\ref{PathMetric}) can be rewritten as
\begin{equation}
M = \sum\limits_{i = 0}^{n-1} \textrm{min}\left\lbrace0,y_i^m\right\rbrace.
\label{PathMetricZero}
\end{equation}

Let $F$  be the cumulative distribution function (CDF) of the normal distribution with mean $2/\sigma^2$ and variance $4/\sigma^2$. Then each element of sum (\ref{PathMetricZero}) is sampled from a random variable with the CDF
\begin{equation}
\tilde{F}\left(x\right) = 
\begin{cases}
    F\left(x\right) ,& \text{if } x < 0\\
    1, & \text{otherwise}.
\end{cases}
\label{TruncPDF}
\end{equation}

Using (\ref{TruncPDF}) and central limit theorem, it is possible to approximate the CDF of the sum (\ref{PathMetricZero}) by the CDF of normal distribution with mean $n\tilde{\mu}$ and variance $n\tilde{\sigma}^2$, where $\tilde{\mu}$ and $\tilde{\sigma}^2$ are mean and variance of a random variable with the CDF given by (\ref{TruncPDF}) correspondingly. Based on the distribution, one can estimate a metric threshold, which will be exceeded during SC decoding with a small probability. The precise value of the threshold can be evaluated in a recursive manner using the following theorem. 

\begin{theorem} 
Let $\tilde{X}$ be a random variable with the CDF $\tilde{F}$ defined by (\ref{TruncPDF}). Let $\tilde{f}\left(x\right)$ be the PDF of $\tilde{X}$ defined on the interval $\left(-\infty, 0\right)$. Let $\tilde{F_n}$ denotes the CDF of the sum of $n$ independent and identically distributed random variables $\tilde{X}$.
Then 
\begin{equation}
\tilde{F}_n(z) = 
 \begin{cases}
 \begin{aligned}
    & \int\limits_{\rightarrow z}^{\rightarrow 0}\tilde{f}_a(x) \tilde{F}_b(z-x) dx\\ 
    & {}+ \tilde{F}_a(z)+ (1-\tilde{F}_a(0-))\tilde{F}_b(z) 
    \end{aligned}, &\text{$z<0$}\\
   	1, & \text{otherwise},
 \end{cases}
 \label{PreciseCDF}
\end{equation}
where $\int\limits_{\rightarrow z}^{\rightarrow 0}f(x)dx = \lim\limits_{u \to z^+}\lim\limits_{v \to 0^-}\int\limits_u^vf(x)dx$, $\tilde{f}_a\left(x\right)$ is $a$-fold convolution of function $\tilde{f}\left(x\right)$ with itself, $a$ and $b$ are some positive integers such that $n=a+b$. 
\label{preciseCDFTheorem}
\end{theorem}

The proof of the theorem is given in the appendix.

%Unfortunately, it has been observed that the threshold value obtained using central limit theorem is greater than the true one. Therefore, a precise CDF of the sum (\ref{PathMetricZero}) has been evaluated and is given by
%\begin{equation}
%\tilde{F}_n(z) = 
% \begin{cases}
% \begin{aligned}
%    & \int\limits_{\rightarrow z}^{\rightarrow 0}\tilde{f}_a(x) \tilde{F}_b(z-x) dx+ \\ 
%    & \tilde{F}_a(z)+ (1-\tilde{F}_a(0-))\tilde{F}_b(z) 
%    \end{aligned}, &\text{$z<0$}\\
%   	1, & \text{otherwise}
% \end{cases},
%\label{PreciseCDF}
%\end{equation}
%where $\int\limits_{\rightarrow z}^{\rightarrow 0}f(x)dx = \lim\limits_{u \to z^+}\lim\limits_{v \to 0^-}\int\limits_u^vf(x)dx$, $\tilde{f}_a(x), \tilde{F}_a(x)$ are a PDF and a CDF of the sum of $a$ random variables defined as (\ref{TruncPDF}), $a$ and $b$ are some natural numbers such that $n=a+b$. 
%% The prove of (\ref{PreciseCDF}) is given in the appendix.

\textit{Example:} Consider that a codeword of length $512$ is transmitted over a BI-AWGN channel with noise variance $\sigma^2 = 0.5$. Then the PDF of the sum (\ref{PathMetricZero}) can be approximated by the normal distribution with mean $-50.95$ and variance $106.1$. Let the threshold value be exceeded with probability $10^{-4}$. Then the threshold equals $F^{-1}\left(10^{-4}\right) = -89.77$, where $F$ is the CDF of the normal distribution, used for the approximation. If (\ref{PreciseCDF}) is used, then the threshold value equals $-96.68$. Note that the value obtained using central limit theorem is greater than the value evaluated using Theorem \ref{preciseCDFTheorem}.

The formal description of the proposed permutation decoding method with both early termination techniques is given in Figs. \ref{RecursivelyCalcSC} -- \ref{PermutationDecodingAlgorithm}. Our modification of the SC decoding algorithm supposes that frozen bits equal zero. It also evaluates metric (\ref{PathMetricSCPerm}). It is assumed that the branch and bounds method have no information about the metric threshold $M_t$. In contrast, SNR based approach improves the first early termination method by setting $M_t$ before the algorithm starts. If the instances of the SC decoder run in parallel, then the threshold value can be used to decrease computational complexity. Note that the block error rate (BLER) of the permutation decoder
will be lower bounded by the probability that has been used to evaluate a threshold value. Also, it is possible to use both techniques together.

\subsection{Repetition handling approach}
It has been observed that the correct codeword can be returned by several instances of the SC decoder used in the considered permutation decoding method. So, if a codeword is returned by $L_c$ copies of the SC decoder and it has the best metric found so far, then it is proposed to stop the decoding procedure and return the codeword found.

 \begin{figure}[tbp]
 \begin{algorithmic}[1]
 \renewcommand{\algorithmicrequire}{\textbf{Input:}}
 \renewcommand{\algorithmicensure}{\textbf{Output:}}
 \Require A code length $n$, a set of the frozen bits $\mathcal{F}$, a vector of received channel LLRs $\mathbf{\hat{y}}$, a set of permutations $\mathcal{P}$ of size $L$, a  metric threshold $M_t$.
 \Ensure  A vector of decoded bits $\mathbf{\hat{u}}$, a decoded codeword metric $M$.
\Function{PermDecoding}{$n,\mathcal{F}, \mathbf{\hat{y}}, \mathcal{P}, L, M_t$}
 \State $M \gets M_t$
 \State $m = \log_{2}{n}$
 \State Set $\mathbf{\hat{u}}$ to be all zeros vector of size $n$
 \ForAll {$\pi \in \mathcal{P}$}
 \State Set $\mathbf{\hat{u}^0}$ to be all zeros vector of size $n$
 \State $M^\prime, \mathbf{\hat{u}^m} \gets \texttt{SC}\left(\pi\left(\mathbf{\hat{y}}\right), \mathbf{\hat{u}^0}, \mathcal{F},  0, m, M_t\right)$
 \If {$M^\prime > M$}
 \State $M \gets M^\prime$
 \State $\mathbf{\hat{u}} \gets \pi^{-1}\left(\mathbf{\hat{u}^0}\right)$
 \EndIf
 \EndFor
 \Return $\mathbf{\hat{u}}, M$ 
 \EndFunction
 \end{algorithmic} 
 \caption{The permutation decoding algorithm.}
 \label{PermutationDecodingAlgorithm}
 \end{figure}

It is an open question, how to compute $L_c$ for a given list size $L$. On the one hand, if $L_c$ is too small, then early termination gain will be enormous, but the error correction performance of the decoder can be degraded. On the other hand, if $L_c$ is too large, then there will be no performance degradation, but the early termination gain will also be negligible. Here simulations are used to determine $L_c$ value. $L_c = 8$ demonstrates almost no error correction performance degradation for $L = 256$, with the early termination gain being significant. Note that the approach cannot be used for a parallel implementation of the permutation decoding algorithm.

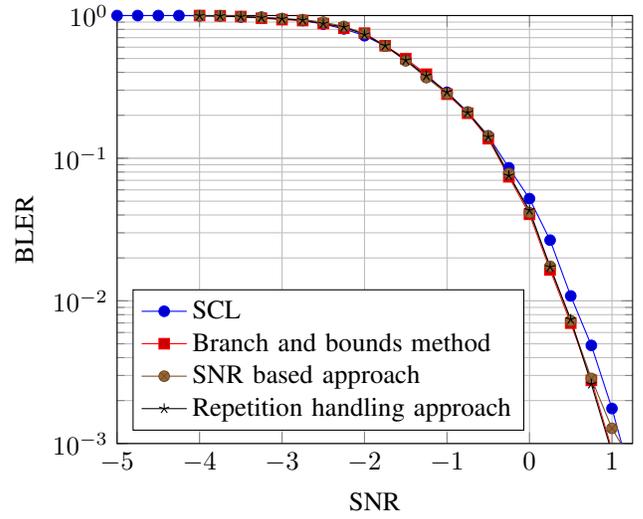
\begin{figure}[tbp]
\centering
\begin{tikzpicture}[ scale = 1.0 ]
	\begin{semilogyaxis}[xlabel=SNR, ylabel=BLER, xmin=-5, xmax=1.25, ymax=1, ymin=0.001, legend entries={SCL, Branch and bounds method, SNR based approach, Repetition handling approach}, grid=major, yminorgrids=true,legend pos=south west, legend cell align=left]	
	\addplot coordinates {
(-5.000000,1.000000)
(-4.750000,1.000000)
(-4.500000,1.000000)
(-4.250000,1.000000)
(-4.000000,0.996016)
(-3.750000,0.996016)
(-3.500000,0.976562)
(-3.250000,0.972763)
(-3.000000,0.950570)
(-2.750000,0.922509)
(-2.500000,0.871080)
(-2.250000,0.806452)
(-2.000000,0.722543)
(-1.750000,0.611247)
(-1.500000,0.491159)
(-1.250000,0.370370)
(-1.000000,0.289687)
(-0.750000,0.210438)
(-0.500000,0.143349)
(-0.250000,0.085704)
(0.000000,0.052062)
(0.250000,0.026644)
(0.500000,0.010823)
(0.750000,0.004871)
(1.000000,0.001759)
(1.250000,0.000524)
	};	
	\addplot coordinates {
		(-4.000000,1.000000)
(-3.750000,0.996016)
(-3.500000,0.984252)
(-3.250000,0.957854)
(-3.000000,0.936330)
(-2.750000,0.922509)
(-2.500000,0.880282)
(-2.250000,0.816993)
(-2.000000,0.753012)
(-1.750000,0.614251)
(-1.500000,0.499002)
(-1.250000,0.388199)
(-1.000000,0.281215)
(-0.750000,0.206782)
(-0.500000,0.137061)
(-0.250000,0.074118)
(0.000000,0.040460)
(0.250000,0.016476)
(0.500000,0.006979)
(0.750000,0.002774)
(1.000000,0.000902)
	};
	\addplot coordinates {
		(-4.000000,0.996016)
(-3.750000,0.996016)
(-3.500000,0.984252)
(-3.250000,0.972763)
(-3.000000,0.950570)
(-2.750000,0.936330)
(-2.500000,0.899281)
(-2.250000,0.838926)
(-2.000000,0.748503)
(-1.750000,0.611247)
(-1.500000,0.481696)
(-1.250000,0.367647)
(-1.000000,0.283768)
(-0.750000,0.210438)
(-0.500000,0.143184)
(-0.250000,0.078101)
(0.000000,0.041514)
(0.250000,0.017470)
(0.500000,0.007255)
(0.750000,0.002859)
(1.000000,0.001277)
(1.250000,0.000723)
	};
	\addplot coordinates {
		(-4.000000,0.996016)
(-3.750000,0.992063)
(-3.500000,0.980392)
(-3.250000,0.965251)
(-3.000000,0.946970)
(-2.750000,0.929368)
(-2.500000,0.877193)
(-2.250000,0.830565)
(-2.000000,0.735294)
(-1.750000,0.615764)
(-1.500000,0.485437)
(-1.250000,0.378788)
(-1.000000,0.290698)
(-0.750000,0.208681)
(-0.500000,0.140924)
(-0.250000,0.075506)
(0.000000,0.043290)
(0.250000,0.017253)
(0.500000,0.007429)
(0.750000,0.002606)
(1.000000,0.000860)
	};
	\end{semilogyaxis}
	\end{tikzpicture}
	\caption{The error correction performance of the RM code of dimension 93 and length 256 under SCL decoding and the proposed permutation decoder. List size equals 256.}
\label{PermVsSCL}
\end{figure}

\begin{figure}[tbp]
\centering
\begin{tikzpicture}[ scale = 1.0 ]
	\begin{axis}[xlabel=SNR, ylabel=Early termination gain, xmin=-10, xmax=7, ymax=2, ymin=1, legend entries={RM $k = 37$ $n = 256$, RM $k = 93$ $n = 256$, RM $k = 163$ $n = 256$, RM $k = 219$ $n = 256$}, grid=major, yminorgrids=true,legend pos=north west, legend cell align=left]	
	\addplot coordinates {
(-13.000000,1.036429)
(-12.750000,1.037126)
(-12.500000,1.037519)
(-12.250000,1.037931)
(-12.000000,1.038268)
(-11.750000,1.038690)
(-11.500000,1.039075)
(-11.250000,1.039476)
(-11.000000,1.039777)
(-10.750000,1.040074)
(-10.500000,1.040457)
(-10.250000,1.040977)
(-10.000000,1.041573)
(-9.750000,1.042335)
(-9.500000,1.042868)
(-9.250000,1.043589)
(-9.000000,1.044901)
(-8.750000,1.045895)
(-8.500000,1.047337)
(-8.250000,1.049315)
(-8.000000,1.051655)
(-7.750000,1.054136)
(-7.500000,1.057214)
(-7.250000,1.059981)
(-7.000000,1.063680)
(-6.750000,1.068413)
(-6.500000,1.072440)
(-6.250000,1.077736)
(-6.000000,1.084416)
(-5.750000,1.090377)
(-5.500000,1.095702)
(-5.250000,1.100488)
(-5.000000,1.104823)
(-4.750000,1.108856)
(-4.500000,1.111506)
(-4.250000,1.112477)
(-4.000000,1.111260)
(-3.750000,1.107918)
(-3.500000,1.103012)
(-3.250000,1.096180)
	};	
	\addplot coordinates {
		(-4.000000,1.130475)
(-3.750000,1.131583)
(-3.500000,1.133947)
(-3.250000,1.137584)
(-3.000000,1.139298)
(-2.750000,1.145037)
(-2.500000,1.152662)
(-2.250000,1.162914)
(-2.000000,1.176316)
(-1.750000,1.200302)
(-1.500000,1.228291)
(-1.250000,1.258393)
(-1.000000,1.288653)
(-0.750000,1.317630)
(-0.500000,1.344099)
(-0.250000,1.365713)
(0.000000,1.375574)
(0.250000,1.374108)
(0.500000,1.354912)
(0.750000,1.323668)
(1.000000,1.281385)
	};
	\addplot coordinates {
		(0.000000,1.327099)
(0.250000,1.327038)
(0.500000,1.331376)
(0.750000,1.336589)
(1.000000,1.347222)
(1.250000,1.364762)
(1.500000,1.398969)
(1.750000,1.434043)
(2.000000,1.489725)
(2.250000,1.555506)
(2.500000,1.621454)
(2.750000,1.678325)
(3.000000,1.705319)
(3.250000,1.678158)
(3.500000,1.595244)
(3.750000,1.483463)
(4.000000,1.369425)
(4.250000,1.265176)
	};
	\addplot coordinates {
		(3.000000,1.715311)
(3.250000,1.715402)
(3.500000,1.721063)
(3.750000,1.729853)
(4.000000,1.755810)
(4.250000,1.785205)
(4.500000,1.819632)
(4.750000,1.868586)
(5.000000,1.863283)
(5.250000,1.804823)
(5.500000,1.700965)
(5.750000,1.560943)
(6.000000,1.421458)
(6.250000,1.294770)
(6.500000,1.193057)
(6.750000,1.116847)
(7.000000,1.066396)
	};
	\end{axis}
	\end{tikzpicture}
\caption{The early termination gain of the branch and bounds method.}
\label{ET1}
\end{figure}
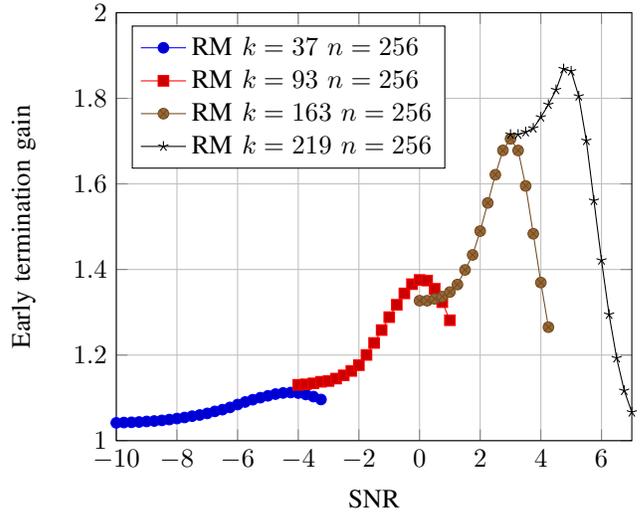

\section{Simulation results}
RM codes of length 256 and different orders are considered. It is assumed that transmission is performed over BI-AWGN channel. All simulations are performed till BLER $10^{-3}$. The comparison of the SCL decoder with the proposed permutation decoding approach is presented in Fig. \ref{PermVsSCL}. The list of size 256 is used for both decoders. Since list size is quite large, the repetition handling approach stops the decoding procedure after eight identical codewords are returned by different copies of the SC decoder. It guarantees that decoding will not stop before the correct codeword is found. The metric threshold has been evaluated using (\ref{PreciseCDF}) as $\tilde{F}^{-1}_{256}(5 \cdot 10^{-4})$. It can be seen that the considered permutation decoding method has a similar error correction performance to that of the SCL algorithm. Moreover, the early termination techniques have a negligible effect on the decoder performance.

\begin{figure}[tbp]
\centering
\begin{tikzpicture}[ scale = 1.0 ]
	\begin{axis}[xlabel=SNR, ylabel=Early termination gain, xmin=-10, xmax=7, ymax=1.25, ymin=1, legend entries={RM $k = 37$ $n = 256$, RM $k = 93$ $n = 256$, RM $k = 163$ $n = 256$, RM $k = 219$ $n = 256$}, grid=major, yminorgrids=true,legend pos=north west, legend cell align=left]	
	\addplot coordinates {
(-10.000000,1.000000)
(-9.750000,1.000000)
(-9.500000,1.000000)
(-9.250000,1.000000)
(-9.000000,1.000000)
(-8.750000,1.000000)
(-8.500000,1.000000)
(-8.250000,1.000000)
(-8.000000,1.000000)
(-7.750000,1.000000)
(-7.500000,1.000001)
(-7.250000,1.000005)
(-7.000000,1.000017)
(-6.750000,1.000046)
(-6.500000,1.000108)
(-6.250000,1.000274)
(-6.000000,1.000627)
(-5.750000,1.001183)
(-5.500000,1.002066)
(-5.250000,1.003445)
(-5.000000,1.005289)
(-4.750000,1.007686)
(-4.500000,1.010315)
(-4.250000,1.013084)
(-4.000000,1.016087)
(-3.750000,1.018940)
(-3.500000,1.021681)
(-3.250000,1.023961)
(-3.000000,1.025515)
	};	
	\addplot coordinates {
		(-5.000000,1.000000)
(-4.750000,1.000000)
(-4.500000,1.000000)
(-4.250000,1.000000)
(-4.000000,1.000000)
(-3.750000,1.000000)
(-3.500000,1.000000)
(-3.250000,1.000000)
(-3.000000,1.000001)
(-2.750000,1.000018)
(-2.500000,1.000101)
(-2.250000,1.000459)
(-2.000000,1.001632)
(-1.750000,1.004660)
(-1.500000,1.010949)
(-1.250000,1.021620)
(-1.000000,1.036894)
(-0.750000,1.054170)
(-0.500000,1.070316)
(-0.250000,1.086175)
(0.000000,1.100735)
(0.250000,1.111770)
(0.500000,1.118595)
(0.750000,1.120078)
(1.000000,1.115834)
(1.250000,1.106546)
	};
	\addplot coordinates {
		(0.000000,1.000000)
(0.250000,1.000000)
(0.500000,1.000000)
(0.750000,1.000000)
(1.000000,1.000004)
(1.250000,1.000113)
(1.500000,1.001039)
(1.750000,1.005771)
(2.000000,1.019769)
(2.250000,1.048977)
(2.500000,1.090268)
(2.750000,1.132848)
(3.000000,1.168015)
(3.250000,1.187782)
(3.500000,1.188627)
(3.750000,1.181763)
(4.000000,1.167184)
(4.250000,1.142103)
(4.500000,1.111280)
	};
	\addplot coordinates {
		(3.000000,1.000000)
(3.250000,1.000000)
(3.500000,1.000000)
(3.750000,1.000000)
(4.000000,1.000000)
(4.250000,1.000006)
(4.500000,1.000142)
(4.750000,1.001534)
(5.000000,1.008274)
(5.250000,1.026670)
(5.500000,1.055155)
(5.750000,1.082555)
(6.000000,1.099614)
(6.250000,1.100307)
(6.500000,1.088245)
(6.750000,1.068117)
(7.000000,1.046260)
	};
	\end{axis}
	\end{tikzpicture}
\caption{The early termination gain of the SNR based approach.}
\label{ET2}
\end{figure}
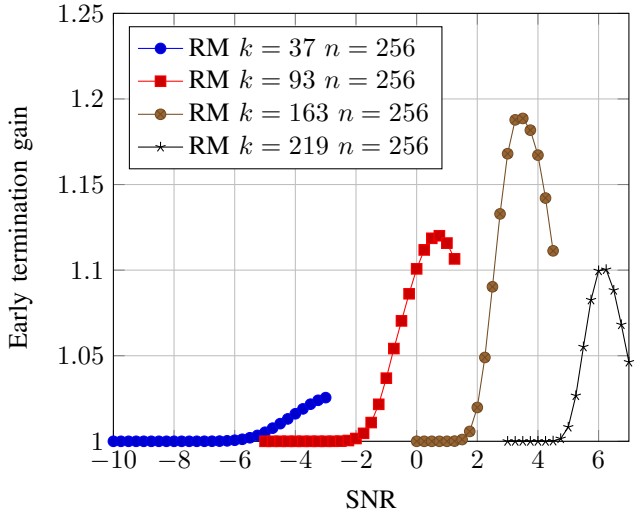

\begin{figure}[tbp]
\centering
\begin{tikzpicture}[ scale = 1.0 ]
	\begin{axis}[xlabel=SNR, ylabel=Early termination gain, xmin=-10, xmax=7, ymax=25, ymin=1, legend entries={RM $k = 37$ $n = 256$, RM $k = 93$ $n = 256$, RM $k = 163$ $n = 256$, RM $k = 219$ $n = 256$}, grid=major, yminorgrids=true,legend pos=north west, legend cell align=left]	
	\addplot coordinates {
(-10.000000,1.034319)
(-9.750000,1.038115)
(-9.500000,1.040618)
(-9.250000,1.039618)
(-9.000000,1.043794)
(-8.750000,1.042680)
(-8.500000,1.053795)
(-8.250000,1.064413)
(-8.000000,1.089680)
(-7.750000,1.105045)
(-7.500000,1.132771)
(-7.250000,1.146556)
(-7.000000,1.187164)
(-6.750000,1.246209)
(-6.500000,1.315812)
(-6.250000,1.396901)
(-6.000000,1.529908)
(-5.750000,1.672919)
(-5.500000,1.880720)
(-5.250000,2.129837)
(-5.000000,2.469632)
(-4.750000,2.896924)
(-4.500000,3.492923)
(-4.250000,4.277873)
(-4.000000,5.275553)
(-3.750000,6.579350)
(-3.500000,8.036258)
(-3.250000,9.756738)
	};	
	\addplot coordinates {
		(-5.000000,1.000000)
(-4.750000,1.000000)
(-4.500000,1.000000)
(-4.250000,1.000000)
(-4.000000,1.000000)
(-3.750000,1.000000)
(-3.500000,1.000000)
(-3.250000,1.000000)
(-3.000000,1.000000)
(-2.750000,1.001323)
(-2.500000,1.009450)
(-2.250000,1.017725)
(-2.000000,1.031597)
(-1.750000,1.064209)
(-1.500000,1.095726)
(-1.250000,1.167488)
(-1.000000,1.272550)
(-0.750000,1.415409)
(-0.500000,1.641613)
(-0.250000,1.941806)
(0.000000,2.400546)
(0.250000,3.168839)
(0.500000,4.291380)
(0.750000,5.913393)
(1.000000,8.104765)
	};
	\addplot coordinates {
		(0.000000,1.000000)
(0.250000,1.000000)
(0.500000,1.000000)
(0.750000,1.000061)
(1.000000,1.001175)
(1.250000,1.007818)
(1.500000,1.019172)
(1.750000,1.049770)
(2.000000,1.102760)
(2.250000,1.181532)
(2.500000,1.328027)
(2.750000,1.550520)
(3.000000,1.915054)
(3.250000,2.558329)
(3.500000,3.705187)
(3.750000,5.492690)
(4.000000,8.206537)
(4.250000,11.838753)
	};
	\addplot coordinates {
		(3.000000,1.051162)
(3.250000,1.056981)
(3.500000,1.055602)
(3.750000,1.062811)
(4.000000,1.079721)
(4.250000,1.125169)
(4.500000,1.186302)
(4.750000,1.314039)
(5.000000,1.610796)
(5.250000,2.058589)
(5.500000,2.754381)
(5.750000,3.847283)
(6.000000,5.819801)
(6.250000,8.986670)
(6.500000,13.498224)
(6.750000,18.727754)
(7.000000,23.267532)
	};
	\end{axis}
	\end{tikzpicture}
\caption{The early termination gain of the repetition handling approach.}
\label{ET3}
\end{figure}
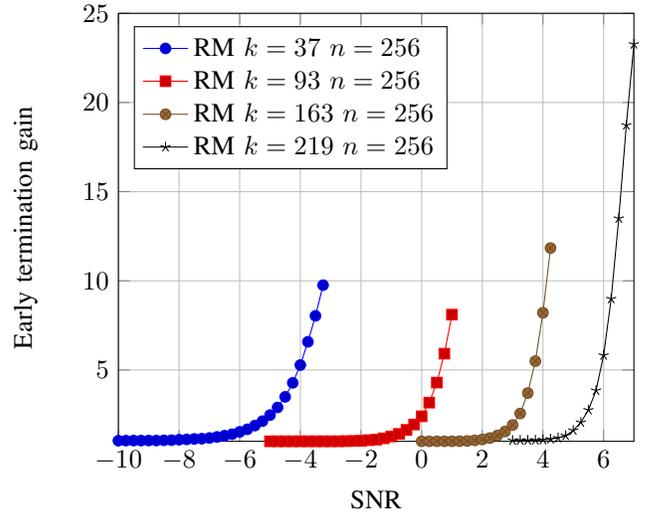

Let $q$ be the number of operations $f_+$ and $\tilde{f_{-}}$ required by the original version of the proposed permutation decoding method and let $q^{et}$ be the number of the same operations after applying an early termination technique. Note that $q$ can be calculated as $q = Ln\log_2n$, where $n$ is the code length and $L$ is the list size, while $q^{et}$ is estimated using simulations. The early termination gain, calculated as $q / q^{et}$, is depicted in Figs. \ref{ET1} -- \ref{ET3}. The largest gain is obtained using repetition handling approach. The permutation decoding algorithm with this early termination technique requires 23 times fewer calculations for the high rate codes in high SNR region in comparison with the original approach.
%The early termination gain can be further increased by accurate choice of a repetitions number. But it can lead to the error correction performance degradation.

The branch and bounds method also demonstrates the best result for the high rate code, with the early termination gain being equal to $1.86$. The number of operations required for decoding with the branch and bounds method is decreasing with the code rate increase.

The worst results are shown by SNR based approach, with the maximum early termination gain being equal to $1.18$. Note that the gain can be further improved by a proper choice of the threshold. For the experiment, the threshold value has been fixed for BLER $5\cdot10^{-4}$, while it can be dynamically chosen based on the decoder performance. This can lead to a greater early termination gain in the cost of the error correction performance. Also, it is the only method that can be used for the fully parallel implementation of considered permutation decoding.

\section{Conclusion}
A new permutation decoding method for RM codes is presented. It has a similar error correction performance to that of the SCL decoding algorithm while having the same complexity. Since sorting operations are not used, it is more feasible for hardware implementation. Also, a parallel implementation of the proposed permutation decoding algorithm is possible, leading to latency improvement. Moreover, it benefits from several early termination techniques, decreasing the number of operations up to 23 times.
 
\section*{Appendix}
\begin{IEEEproof}[\IEEEproofname{ of Theorem \ref{TheoremMetric}}]
Observe that 
\begin{equation*}
\begin{aligned}
\sum\limits_{i = 0}^{n-1} \textrm{min}\left\lbrace0,y^{0}_{i}\left(1-2\hat{u}^{0}_{i}\right) \right\rbrace = \sum_{i \in \mathcal{F}} \textrm{min}\left\lbrace0,y_i^0\left(1 - 2\hat{u}_i^0\right) \right\rbrace.
\end{aligned}
\end{equation*} 
So, to prove (\ref{PathMetric}), one needs to show that  
\begin{equation*}
\begin{aligned}
& \sum\limits_{i = 0}^{n-1} \textrm{min}\left\lbrace0,y^{l}_{i}\left(1-2\hat{u}^{l}_{i}\right) \right\rbrace \\
& = \sum\limits_{i = 0}^{n-1} \textrm{min}\left\lbrace0,y^{l-1}_{i}\left(1-2\hat{u}^{l-1}_{i}\right)\right\rbrace,
\end{aligned}
\end{equation*} 
then the statement of the theorem is followed by recursion. For simplicity, we will show that 
\begin{equation}
\begin{aligned}
& \textrm{min}\left\lbrace0,y_0\left(1-2\left(u_0 \oplus u_1\right)\right)\right\rbrace + 
\textrm{min}\left\lbrace0,y_1\left(1-2u_1\right)\right\rbrace \\
& =  \textrm{min}\left\lbrace0,\tilde{f_{-}}\left(y_0,y_1\right)  \left(1-2u_0\right)\right\rbrace \\
& + \textrm{min}\left\lbrace0,f_{+}\left(y_0,y_1, u_0\right) \left(1-2u_1\right)\right\rbrace,
\end{aligned}
\label{SimpleRec}
\end{equation}
where $u_0, u_1 \in \left\lbrace 0, 1 \right\rbrace$ and $y_0$, $y_1 \in \mathbb{R}$. To show that (\ref{SimpleRec}) holds, we will consider two cases, namely
\begin{equation*}
\begin{aligned}
\textrm{sign}\left(y_0\right)\textrm{sign}\left(y_1\right) & = 1-2u_0 \\
\textrm{sign}\left(y_0\right)\textrm{sign}\left(y_1\right) & = \left(1-2\left(u_0 \oplus 1\right)\right).
\end{aligned}
\end{equation*}

If $$\textrm{sign}\left(y_0\right)\textrm{sign}\left(y_1\right) = 1-2u_0,$$
 then $\textrm{min}\left\lbrace0,\tilde{f_{-}}\left(y_0,y_1\right)\left(1-2u_0\right)\right\rbrace = 0$ and $y_0\left(1-2u_0\right)$, $y_1$ are both positive or negative. Thus, 

\begin{equation*}
\begin{aligned}
& \textrm{min}\left\lbrace0,y_0\left(1-2\left(u_0 \oplus u_1\right)\right)\right\rbrace + 
\textrm{min}\left\lbrace0,y_1\left(1-2u_1\right)\right\rbrace \\
& = \textrm{min}\left\lbrace0,y_0\left(1-2u_0\right)\left(1-2u_1\right)\right\rbrace + 
\textrm{min}\left\lbrace0,y_1\left(1-2u_1\right)\right\rbrace \\
& = \textrm{min}\left\lbrace0,y_0\left(1-2u_0\right)\left(1-2u_1\right) + y_1\left(1-2u_1\right)\right\rbrace \\
& = \textrm{min}\left\lbrace0,y_0\left(1-2\left(u_0 \oplus u_1\right)\right) + y_1\left(1-2u_1\right)\right\rbrace \\
& = \textrm{min}\left\lbrace0,f_{+}\left(y_0,y_1, u_0\right) \left(1-2u_1\right)\right\rbrace
\end{aligned}
\end{equation*}
and the statement holds.

Consider the second case, i.e. $y_0\left(1-2u_0\right)$ and $y_1$ have different signs. To prove it, we need to consider four cases, namely
\begin{equation*}
\begin{aligned}
y_0\left(1-2\left(u_0 \oplus u_1\right)\right) < 0, \left|y_0\right| > \left|y_1\right|, \\
y_0\left(1-2\left(u_0 \oplus u_1\right)\right) < 0, \left|y_0\right| \leq \left|y_1\right|, \\
y_0\left(1-2\left(u_0 \oplus u_1\right)\right) \geq 0, \left|y_0\right| > \left|y_1\right|, \\
y_0\left(1-2\left(u_0 \oplus u_1\right)\right) \geq 0, \left|y_0\right| \leq \left|y_1\right|. \\
\end{aligned}
\end{equation*}
Here we prove the first case. All the others are proved in a similar way.

 Assume that $y_0\left(1-2\left(u_0 \oplus u_1\right)\right) < 0$. Then 
\begin{equation*}
\begin{aligned}
& \textrm{min}\left\lbrace0,y_0\left(1-2\left(u_0 \oplus u_1\right)\right)\right\rbrace + 
\textrm{min}\left\lbrace0,y_1\left(1-2u_1\right)\right\rbrace \\
& = y_0\left(1-2\left(u_0 \oplus u_1\right)\right).
\end{aligned}
\end{equation*}
Let $\left|y_0\right| > \left|y_1\right|$, then 
\begin{equation*}
\begin{aligned}
&\textrm{min}\left\lbrace0,f_{+}\left(y_0,y_1, u_0\right) \left(1-2u_1\right)\right\rbrace \\
& = \textrm{min}\left\lbrace0,y_0\left(1-2\left(u_0 \oplus u_1\right)\right) + y_1\left(1-2u_1\right)\right\rbrace \\
& = y_0\left(1-2\left(u_0 \oplus u_1\right)\right) + y_1\left(1-2u_1\right),
\end{aligned}
\end{equation*}
\begin{equation*}
\begin{aligned}
& \textrm{min}\left\lbrace0,\tilde{f_{-}}\left(y_0,y_1\right)\left(1-2u_0\right)\right\rbrace \\
& = \left|y_1\right|\left(1-2\left(u_0 \oplus 1\right)\right)\left(1-2u_0\right) = -\left|y_1\right|.
\end{aligned}
\end{equation*}

Since $y_1\left(1-2u_1\right) >= 0$, it follows that $y_1\left(1-2u_1\right) = \left|y_1\right|$, and the statement holds.
\end{IEEEproof}

\begin{IEEEproof}[\IEEEproofname{ of Theorem \ref{preciseCDFTheorem}}]
Recall that $\int\limits_{\rightarrow z}^{\rightarrow 0}f(x)dx = \lim\limits_{u \to z^+}\lim\limits_{v \to 0^-}\int\limits_u^vf(x)dx$. To prove Theorem \ref{preciseCDFTheorem} we need the following lemma.
\begin{lemma}
Let $\tilde{f}\left(x\right)$ be the PDF of a normal distribution defined on the interval $\left(-\infty, 0\right)$. Let us define $n$-fold convolution of function $\tilde{f}\left(x\right)$ with itself as $\tilde{f}_n(x)$. Then
\begin{equation*}
\begin{aligned}
\tilde{f}_n(z) =  \int\limits_{\rightarrow z}^{\rightarrow 0} \tilde{f}_a(x)\tilde{f}_b(z-x)dx ,
\end{aligned}
\end{equation*}
 $a$ and $b$ are some positive integers such that $n=a+b$. 
\end{lemma}
\begin{IEEEproof}
Let $\ast$ denotes the convolution operation. Convolution of two functions \cite[eq.~(6.39)]{papoulis} is defined as 
\begin{equation*}
\begin{aligned}
\left(\tilde{f}\ast\tilde{g}\right)\left(z\right) =  \int\limits_{-\infty}^{\infty} \tilde{f}(x)\tilde{g}(z-x)dx,
\end{aligned}
\end{equation*}
but, since the considered function $\tilde{f}(x)$ is defined on the interval $\left(-\infty, 0\right)$, we will use the following modification
\begin{equation*}
\begin{aligned}
\left(\tilde{f}\ast\tilde{g}\right)\left(z\right) =  \int\limits_{\rightarrow z}^{\rightarrow 0} \tilde{f}(x)\tilde{g}(z-x)dx.
\end{aligned}
\end{equation*}

To prove the statement of the lemma ones require to show that the considered convolution operation is associative. Then
\begin{equation*}
\begin{aligned}
& \left(\left(\tilde{f}*\tilde{g}\right)\ast\tilde{h}\right)\left(t\right) \\
& = \int\limits_{\rightarrow t}^{\rightarrow 0} \left(\tilde{f}*\tilde{g}\right)(x)\tilde{h}(t-x)dx \\
& = \int\limits_{s=\rightarrow t}^{\rightarrow 0} \left(\int\limits_{u =\rightarrow s}^{\rightarrow 0} \tilde{f}\left(u\right)\tilde{g}\left(s-u\right)du\right)h\left(t-s\right)ds \\
& = \int\limits_{s=\rightarrow t}^{\rightarrow 0} \int\limits_{u =\rightarrow s}^{\rightarrow 0} \tilde{f}\left(u\right)\tilde{g}\left(s-u\right)h\left(t-s\right)duds \\
& = \int\limits_{u=\rightarrow t}^{\rightarrow 0} \int\limits_{s =\rightarrow t}^{\rightarrow u} \tilde{f}\left(u\right)\tilde{g}\left(s-u\right)h\left(t-s\right)dsdu \\
& = \int\limits_{u=\rightarrow t}^{\rightarrow 0} \int\limits_{s =\rightarrow t-u}^{\rightarrow 0} \tilde{f}\left(u\right)\tilde{g}\left(s\right)h\left(t-s-u\right)dsdu \\
& = \int\limits_{u=\rightarrow t}^{\rightarrow 0} \tilde{f}\left(u\right) \left(\int\limits_{s =\rightarrow t-u}^{\rightarrow 0} \tilde{g}\left(s\right)h\left(t-s-u\right)ds\right)du \\
& = \int\limits_{u=\rightarrow t}^{\rightarrow 0} \tilde{f}\left(u\right) \left(g\ast h\right)\left(t-u\right)du =  
\left(\tilde{f}*\left(\tilde{g}\ast\tilde{h}\right)\right)\left(t\right).&
\end{aligned}
\end{equation*}

Using this property, it is easy to see that
\[
\newcommand{\mz}
{ \textcolor{blue}{ \scriptstyle{ - 1 } } }
\newcommand{\pz}{ \textcolor{red} 
{\scriptstyle{\hphantom{-} 1}}}
\begin{aligned}
\tilde{f}_n\left(z\right) & = \left(\underbrace{\left(\tilde{f}*\tilde{f}*\dots*\tilde{f}\right)}_{n-1}*\tilde{f}\right)\left(z\right) \\
& = \left(\underbrace{\left(\tilde{f}*\tilde{f}*\dots*\tilde{f}\right)}_{n-2}*\tilde{f}*\tilde{f}\right)\left(z\right)  \\
& = \left(\underbrace{\tilde{f}*\tilde{f}*\dots*\tilde{f}}_{a}*\underbrace{\tilde{f}*\tilde{f}*\dots*\tilde{f}}_{b}\right)\left(z\right)\\
& = \left(\tilde{f}_{a}*\tilde{f}_{b}\right)\left(z\right)  = \int\limits_{\rightarrow z}^{\rightarrow 0} \tilde{f}_{a}(x)\tilde{f}_{b}(z-x)dx.
\end{aligned}
%\underbrace{\overbrace{ 
%\begin{array} 
%{@{}r@{\,}r@{\,}r@{\,}r@{}}
%1 & 0 & \dots & 0 \\
%\mz & \mz & \dots & \mz
%\end{array}
%}^{d+1 \leq j \leq k+1}
%\underbrace{\begin{array}{@{}r@{\,}r@{\,}r@{}}
%1 & 0 & \dots \\
%\pz & \pz & \dots
%\end{array}
%}_{\trigntjm}
%}_{\trigntjm = \sum_{t=0}^{j-1}(-1)\prou^{-mt}
%+\phase \trigntjm 
%\displaystyle . 
%}
\]
\end{IEEEproof}

Let $z \geq 0$. Observe that $\tilde{F}\left(0\right) = 1$ Then $\tilde{F}_n\left(0\right) = 1$. Since $\tilde{F}_n\left(x\right) \leq \tilde{F}_n\left(x + \epsilon \right), \epsilon > 0$, if follows that $\tilde{F}\left(z\right) = 1, z \geq 0$

Consider the case $z < 0$. Let $f$, $g$ be the PDF of random variables $X$ and $Y$ respectively. Then the CDF of the sum of two random variables $X$ and $Y$ \cite[eq.~(6.37)]{papoulis} is defined as 
\begin{equation*}
F\left(z\right) = \int\int_{x + y \leq z} f\left(x\right)g\left(y\right)dxdy.
\end{equation*}
Since a random variable defined by the considered CDF is of mixed type, we need to consider two cases. First, when both random variables take negative values. Second, when one of the random variables is negative, while another equals zero.
Let $\tilde{f}_n(x)$ be $n$-fold convolution of function $\tilde{f}\left(x\right)$ with itself and let
\begin{equation*}
t_{ab}\left(z\right) = \tilde{F}_b(z)\left(1 - \tilde{F}_a(0-) \right) + \tilde{F}_a(z)\left(1 - \tilde{F}_b(0-) \right).
\end{equation*}

Then
\begin{align*}
\tilde{F}_n(z) & = \int\limits_{-\infty}^{z} \tilde{f}_n(x)dx + t_{ab}\left(z\right) \\
& = \int\limits_{y = -\infty}^{z}\left( \int\limits_{x = \rightarrow y}^{\rightarrow 0} \tilde{f}_a(x)\tilde{f}_b(y-x)dx\right)dy + t_{ab}\left(z\right)\\
& = \int\limits_{y = -\infty}^{z}\int\limits_{x = \rightarrow y}^{\rightarrow 0}  \tilde{f}_a(x)\tilde{f}_b(y-x)dxdy + t_{ab}\left(z\right)  \\
& = \int\limits_{y = -\infty}^{z}\int\limits_{x = \rightarrow z}^{\rightarrow 0}  \tilde{f}_a(x)\tilde{f}_b(y-x)dxdy  \\ 
& + \int\limits_{y = -\infty}^{z}\lim\limits_{u \rightarrow y^+}\lim\limits_{v \rightarrow z^+}\int\limits_{x = u}^{v}  \tilde{f}_a(x)\tilde{f}_b(y-x)dxdy + t_{ab}\left(z\right) \\
& = \int\limits_{x = \rightarrow z}^{\rightarrow 0}\int\limits_{y = -\infty}^{z-x}  \tilde{f}_a(x)\tilde{f}_b(y)dydx  \\
& +  \lim\limits_{v \rightarrow z+}\int\limits_{x = -\infty}^{v} \int\limits_{y = -\infty}^{\rightarrow 0} \tilde{f}_a(x)\tilde{f}_b(y)dydx + t_{ab}\left(z\right)  \\
& = \int\limits_{\rightarrow z}^{\rightarrow 0}\tilde{F}_b\left(z-x\right)\tilde{f}_a(x)dx \\ 
& + \lim\limits_{v \rightarrow z+}\int\limits_{x = -\infty}^{v} \tilde{F}_b\left(0-\right) \tilde{f}_a(x)dx + t_{ab}\left(z\right) \\
& = \int\limits_{\rightarrow z}^{\rightarrow 0}\tilde{F}_b\left(z-x\right)\tilde{f}_a(x)dx +
\tilde{F}_a\left(z\right)\tilde{F}_b\left(0-\right) + t_{ab}\left(z\right) \\
& = \int\limits_{\rightarrow z}^{\rightarrow 0}\tilde{F}_b\left(z-x\right)\tilde{f}_a(x)dx + \tilde{F}_b(z)\left(1 - \tilde{F}_a(0-) \right) \\
& + \tilde{F}_a(z)
\end{align*}
and this concludes the proof of Theorem \ref{preciseCDFTheorem}.
\end{IEEEproof}
\bibliographystyle{IEEEtran}
\bibliography{IEEEabrv,myBib}

% Generated by IEEEtran.bst, version: 1.14 (2015/08/26)
\begin{thebibliography}{10}
\providecommand{\url}[1]{#1}
\csname url@samestyle\endcsname
\providecommand{\newblock}{\relax}
\providecommand{\bibinfo}[2]{#2}
\providecommand{\BIBentrySTDinterwordspacing}{\spaceskip=0pt\relax}
\providecommand{\BIBentryALTinterwordstretchfactor}{4}
\providecommand{\BIBentryALTinterwordspacing}{\spaceskip=\fontdimen2\font plus
\BIBentryALTinterwordstretchfactor\fontdimen3\font minus
  \fontdimen4\font\relax}
\providecommand{\BIBforeignlanguage}[2]{{%
\expandafter\ifx\csname l@#1\endcsname\relax
\typeout{** WARNING: IEEEtran.bst: No hyphenation pattern has been}%
\typeout{** loaded for the language `#1'. Using the pattern for}%
\typeout{** the default language instead.}%
\else
\language=\csname l@#1\endcsname
\fi
#2}}
\providecommand{\BIBdecl}{\relax}
\BIBdecl

\bibitem{Muller}
D.~E. Muller, ``Application of boolean algebra to switching circuit design and
  to error detection,'' \emph{Transactions of the I.R.E. Professional Group on
  Electronic Computers}, vol. EC-3, no.~3, pp. 6--12, Sep. 1954.

\bibitem{Reed}
I.~Reed, ``A class of multiple-error-correcting codes and the decoding
  scheme,'' \emph{Transactions of the IRE Professional Group on Information
  Theory}, vol.~4, no.~4, pp. 38--49, Sep. 1954.

\bibitem{Urbanke}
S.~Kudekar, S.~Kumar, M.~Mondelli, H.~D. Pfister, E.~Şaşoǧlu, and R.~L.
  Urbanke, ``Reed–muller codes achieve capacity on erasure channels,''
  \emph{IEEE Transactions on Information Theory}, vol.~63, no.~7, pp.
  4298--4316, July 2017.

\bibitem{Dumer}
I.~Dumer and K.~Shabunov, ``Soft-decision decoding of reed-muller codes:
  recursive lists,'' \emph{IEEE Transactions on Information Theory}, vol.~52,
  no.~3, pp. 1260--1266, March 2006.

\bibitem{Polar}
E.~Arikan, ``Channel polarization: A method for constructing capacity-achieving
  codes for symmetric binary-input memoryless channels,'' \emph{IEEE
  Transactions on Information Theory}, vol.~55, no.~7, pp. 3051--3073, July
  2009.

\bibitem{SCL}
I.~Tal and A.~Vardy, ``List decoding of polar codes,'' \emph{IEEE Transactions
  on Information Theory}, vol.~61, no.~5, pp. 2213--2226, May 2015.

\bibitem{GA}
P.~Trifonov, ``Efficient design and decoding of polar codes,'' \emph{IEEE
  Transactions on Communications}, vol.~60, no.~11, pp. 3221--3227, November
  2012.

\bibitem{MinSum}
A.~Balatsoukas-Stimming, M.~B. Parizi, and A.~Burg, ``{LLR}-based successive
  cancellation list decoding of polar codes,'' \emph{IEEE Transactions on
  Signal Processing}, vol.~63, no.~19, pp. 5165--5179, Oct 2015.

\bibitem{Sloan}
F.~J. MacWilliams and N.~J.~A. Sloane, \emph{The theory of error-correcting
  codes}.\hskip 1em plus 0.5em minus 0.4em\relax Amsterdam, The Netherlands:
  North-Holland, 1977.

\bibitem{Perm}
\BIBentryALTinterwordspacing
N.~Doan, S.~A. Hashemi, M.~Mondelli, and W.~J. Gross, ``On the decoding of
  polar codes on permuted factor graphs.'' [Online]. Available:
  \url{https://arxiv.org/abs/1806.11195v1}
\BIBentrySTDinterwordspacing

\bibitem{Sarah}
S.~J. Johnson, \emph{Iterative error correction: Turbo, low-density
  parity-check and repeat-accumulate codes}.\hskip 1em plus 0.5em minus
  0.4em\relax Cambridge university press, 2009.

\bibitem{papoulis}
A.~Papoulis, \emph{Probability, random variables, and stochastic
  processes}.\hskip 1em plus 0.5em minus 0.4em\relax McGraw Hill, 1991.

\end{thebibliography}

\end{document}